\documentclass[12pt]{article}

\usepackage{amsmath}
\usepackage{amsthm}
\usepackage{amssymb}
\usepackage{xypic}
\xyoption{curve}
\usepackage{tikz}
\usetikzlibrary{shapes,decorations.pathmorphing}
\usetikzlibrary{intersections}
\usetikzlibrary{decorations.markings}
\usepackage{caption}
\usepackage{subcaption}
\definecolor{myurlcolor}{rgb}{0,0,0.7}

\usepackage{hyperref}
\hypersetup{colorlinks,
linkcolor=myurlcolor,
citecolor=myurlcolor,
urlcolor=myurlcolor}

\usepackage[margin=3cm]{geometry}

\newcommand{\maps}{\colon}
\newcommand{\To}{\Rightarrow}
\newcommand{\Mod}{\mathbf{C}}
\newcommand{\Rep}{\mathrm{Rep}}
\newcommand{\Vect}{\mathbf{Vect}}
\newcommand{\C}{\mathbb{C}}
\newcommand{\g}{\mathfrak{g}}
\newcommand{\U}{\mathrm{U}}

\begin{document}

\title{A Characterization of Entropy in Terms of Information Loss} 

%\author{John C. Baez}
%\address{Department of Mathematics\\ 
%University of California\\ 
%Riverside CA 92521\\
%and Centre for Quantum Technologies\\ 
%National University of Singapore\\ 
%Singapore 117543}
%\email{baez@math.ucr.edu}
%\author{John Baez and Jamie Vicary}
%\address{Department of Computer Science \\
%University of Oxford \\
%Oxford OX1 3QD, UK \\
%and Centre for Quantum Technologies, \\
%National University of Singapore \\
%Singapore 117543 }
%\email{jamie.vicary@cs.ox.ac.uk }

\author{\begin{tabular}{c@{\hspace{2cm}}c}
John C.\ Baez\footnote{Department of Mathematics, University of California, Riverside CA, 92521, USA}{ }\footnote{Centre for Quantum Technologies, National University of Singapore, 117543, Singapore} & Jamie Vicary$^{\dag}$\footnote{Department of Computer Science, University of Oxford, Parks Road, Oxford, OX1 3QP, UK}
\\
baez@math.ucr.edu & jamie.vicary@cs.ox.ac.uk
\end{tabular}}

\date{\today}

\title{\bf Wormholes and Entanglement}

\maketitle

\begin{abstract}
\noindent
Maldacena and Susskind have proposed a correspondence between wormholes and entanglement, dubbed ER=EPR. We study this in the context of 3d topological quantum field theory, where we show that the formation of a wormhole is the same process as creating a particle-antiparticle pair. A key feature of the ER=EPR proposal is that certain apparently entangled degrees of freedom turn out to be the same. We name this phenomenon `fake entanglement', and show how it arises in our topological quantum field theory model.
\end{abstract}

\section{Introduction}

A surprising connection between wormholes and entanglement was recently suggested by Maldacena and Susskind~\cite{erepr}. They proposed that whenever two systems are maximally entangled, there will always exist a wormhole with one of the systems at each end. Given the early work of Einstein, Podolsky and Rosen~\cite{epr} on entangled states, and the description by Einstein and Rosen~\cite{er} of wormhole solutions to general relativity, they labelled their proposal `ER=EPR'. Their motivation was to resolve an apparent paradox in the theory of quantum black hole---the `firewall paradox'---which was brought to widespread attention by Almheiri, Marolf, Polchinski and Sully~\cite{amps1,amps2} (see also the references therein).

Inspired by Maldacena and Susskind's proposal, our goal in this paper is to study the relation between wormholes and entanglement in 3-dimensional topological quantum field theory (TQFT). This is a theory in which space is a compact 2-dimensional manifold, possibly with boundary, and spacetime is a compact 3\-dimensional manifold, possibly with boundary and corners. This is perhaps the simplest setting in which wormholes can be studied, since sensitivity at least to topology is required for a wormhole to be discernible, and at least 2 dimensions of space are required for wormholes to exist between points on a connected manifold. Our main result is that the process of creating a particle-antiparticle pair in this theory is identical to the process of creating a wormhole.

\begin{figure}[h]
\centering
\begin{subfigure}[b]{0.3\textwidth}
$\begin{aligned}
\begin{tikzpicture}[thick]
\foreach \x in {0cm,3cm}
{
\begin{scope}[xshift=\x]
\draw [black!20] (0,0) arc (0:180:1 and 0.5);
\draw (0,0) arc (0:-180:1 and 0.5);
\draw (0,3) arc (0:360:1 and 0.5);
\draw (0,0) to (0,3);
\draw (-2,0) to (-2,3);
\draw [black!20] (-0.5,0) arc (0:180:0.5 and 0.25);
\draw (-0.5,0) arc (0:-180:0.5 and 0.25);
\draw (-0.5,3) arc (0:360:0.5 and 0.25);
\draw (-0.5,0) to (-0.5,3);
\draw (-1.5,0) to (-1.5,3);
\draw [decoration={snake,amplitude=0.05cm}, decorate] (-1,0) to (-1,3);
\end{scope}
}
\end{tikzpicture}
\end{aligned}$
\caption{Natural coordinates}
\label{ads1}
\end{subfigure}
\hspace{4cm}
\begin{subfigure}[b]{0.3\textwidth}
$\begin{aligned}
\begin{tikzpicture}[thick]
\draw (0,0) to (0,4);
\draw (4,0) to (4,4);
\draw [decoration={snake}, decorate] (0,0) to (4,0);
\draw [decoration={snake}, decorate] (0,4) to (4,4);
\draw (0,0) to (4,4);
\draw (0,4) to (4,0);
\draw [->] (0.2,0.5) to node [auto, inner sep=1pt, pos=0.7] {$A'$} +(45:0.75cm);
\draw [->] (2.5,2.8) to node [auto, inner sep=1pt] {$A$} +(45:0.75cm);
\draw [->] (2.8,2.5) to node [auto, swap, inner sep=1pt] {$B$} +(45:0.75cm);
\draw [<-] (0.2,3.5) to node [auto, swap, inner sep=1pt, pos=0.85] {$A''$} +(-45:0.75cm);
\draw [dotted] (0.7,1.0) to (2.5,2.8);
\node [circle, draw, inner sep=1pt] at (.7,2) {1};
\node [circle, draw, inner sep=1pt] at (3.3,2) {2};
\end{tikzpicture}
\end{aligned}$
\caption{Penrose diagram}
\label{ads2}
\end{subfigure}
\caption{The extended black hole in anti--de Sitter space}
\end{figure}
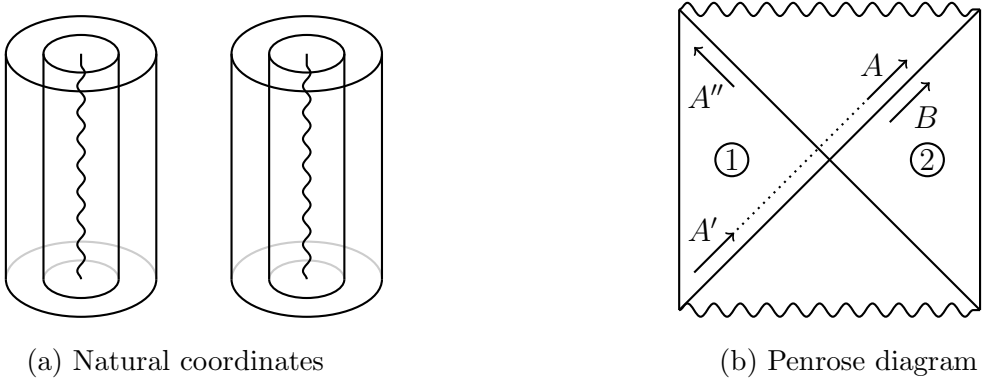

We do not take any position on the firewall paradox, or whether `ER=EPR' can resolve this apparent paradox. However, it will be useful to start with a summary of what Maldacena and Susskind say about wormholes and entanglement.  They make use of the AdS/CFT correspondence, which gives a way to relate quantum gravity on anti--de Sitter spacetime with conformal field theory on its boundary. They use this to analyze the maximally extended black hole solution to general relativity in an asymptotically anti--de Sitter spacetime.  This solution contains two separate universes linked by a wormhole. We can draw this solution in two different ways, shown in Figure~1.  Figure~\ref{ads1} shows the two disjoint universes, each containing a single black hole, where it is understood that the interiors of the black holes should be identified. Figure~\ref{ads2} gives the Penrose diagram for this solution.

In Figure \ref{ads2} we have indicated some modes of the quantum field. Modes $A$ and $B$ represent two entangled particles that have been created in the neighbourhood of the horizon in universe 2. By considering the Penrose diagram, it is clear that that mode $A$ is obtained by bulk time evolution of a mode $A'$ in universe 1, and so in the Heisenberg picture they are equal elements of the Hilbert space of states. The mode $A'$ can be identified as a boundary excitation of universe 1, and thus gives a state in the dual CFT. Since the two CFTs have independent dynamics, evolving mode $A'$ forward in time as a state of the CFT gives rise to a new boundary excitation $A''$ also localised to universe 1, which again is the same state as $A'$ in the Heisenberg picture. So the modes $A$, $A'$ and $A''$ are all identified.

In this scenario the mode $B$ must be maximally entangled to $A$, as they are created by a pair-creation event at the horizon. Yet $B$ must also be maximally entangled to $A''$, due to an argument of Maldacena and Susskind that it could have been formed from radiation that left the black hole at an early stage.   Superficially this may seem to violate `monogamy of entanglement', which says that one quantum system cannot be maximally entangled to two others~\cite{monog1}.  However, the modes $A$ and $A''$ are identical, so there is no contradiction. 

It is striking that this argument of Maldacena and Susskind relies so little on the detailed physics of the AdS/CFT correspondence.  The key aspect is the identification of certain apparently entangled field modes as being the same. We analyze this phenomenon in Section~\ref{sec:fakeentanglement}, naming it `fake entanglement': by constraining the state space of a composite system to a particular subspace containing entangled states, the system gains the ability to be entangled with another system, in an apparent violation of monogamy of entanglement.

In Section~\ref{sec:paircreation} we describe the pair creation process in topological quantum field theory, and demonstrate that it is equivalent to a wormhole creation process. In Section~\ref{sec:doublecategories} we show how double categories can be used to formalize these ideas.  

\section{Fake entanglement}
\label{sec:fakeentanglement}

Maldacena and Susskind's suggested relationship between wormholes and entanglement in fact involves `fake entanglement': a phenomenon that resembles entanglement, but does not obey the usual law of monogamy of entanglement.  

Given a quantum system with finite-dimensional Hilbert space $A \otimes B$, and a state $\psi \in A \otimes B$, we say $A$ is `completely entangled' with $B$ in the state $\psi$ if 
\begin{equation}
\psi = \frac{1}{\sqrt{\dim(A)}} \sum_i e_i \otimes f_i
\end{equation}
where $e_i$ ranges over an orthonormal basis of $A$ and $f_i$ ranges over some orthonormal set of vectors in $B$ (not necessarily a basis.)  

More generally, suppose we are given a three-part quantum system with a finite-dimensional Hilbert space $A \otimes B \otimes C$.  We say $A$ is `completely entangled' with $B$ in a state $\psi \in A \otimes B \otimes C$, if after tracing out over $C$, $\psi$ gives a pure state in which $A$ is maximally entangled with $B$~\cite{cavalcanti}.  Monogamy of entanglement says that in such a three-part system, if $C$ is completely entangled with $A$, then $C$ cannot be completely entangled with $B$.

However, suppose the two-part system containing $B$ and $C$ is constrained in such a way that not all states in the Hilbert space $B \otimes C$ are physically allowed, but only those in a subspace $H \subseteq B \otimes C$.  Then the possibility of `fake entanglement' arises.  

The simplest case occurs when $B$ and $C$ have the same dimension, we choose orthonormal bases $f_i$ for $B$ and $g_i$ for $C$, and we define $H$ to be the subspace of $B \otimes C$ with basis vectors $ f_i \otimes g_i $.    In this case $B$ and $C$ are not two independent systems; the Hilbert space $B \otimes C$ is really just an auxiliary device for describing a single system with Hilbert space $H$.   

We can map any state of $H$ to a state of $B$, as follows:
$$ p \colon f_i \otimes g_i \mapsto f_i $$
and similarly we can map any state of $H$ to a state of $C$:
$$ q \colon f_i \otimes g_i \mapsto g_i $$
These maps are unitary, so they let us think of the system $H$ as being either the system $B$ or the system $C$: these are, as it were, two `views' of $H$.

We can now consider a state $\psi \in A \otimes H$ in which $A$ is completely entangled with $H$.  We can view $\psi$ as a state $(1 \otimes p)\psi$ of $A \otimes B$, and in this state $A$ is completely entangled with $B$.  We can also view $\psi$ as a state $(1 \otimes q)\psi$ of $A \otimes C$, and in this state $A$ is completely entangled with $C$.  However, this does not violate monogamy of entanglement!

Fake entanglement requires more than the fact that a pair of systems are currently in a state which lies in a particular subspace.  For example, consider a pair of entangled photons produced by the annihilation of an electron and a positron with total momentum zero. The joint state of the photons lies in a `diagonal subspace' of states with total momentum zero.  But it is physically possible for a pair of photons to have a state outside this diagonal subspace, so this form entanglement is not fake.

In Maldacena and Susskind's argument, the system $H$ is a wormhole and the systems $A$ and $B$ are its two ends.  These ends may superficially appear to be two separate particles, but in reality they are just two `views' of the same wormhole.

\section{Creating entangled pairs}
\label{sec:paircreation}

In 3d TQFTs, particles can be described as topological defects.  The process of creating a pair of particles in a 3d TQFT begins with a single disk. This represents a compact region of the surface on which the quantum field is defined, with a boundary that separates it from the rest of the surface. At a point of the disk a new circular internal boundary grows from size zero to some finite size, a process that changes the topology of the region. The quantum field is no longer defined in the area enclosed by this new internal boundary.

In a solid-state model where the TQFT is an effective field theory, as is believed to be achievable using the fractional quantum Hall effect~\cite{Wang}, this internal boundary may enclose a region where an external perturbation has been applied, for example by the introduction of the uncharged tip of a scanning electron microscope. In the immediate vicinity of the tip the topological order is destroyed, but at sufficient distance, outside the new boundary circle, the TQFT is still well-behaved. A measurement of the total topological charge about a circle that encloses the new internal boundary will give zero.

The shape of this internal boundary circle could change over time by pinching across a diameter and undergoing a second topology change, splitting into two circles. Physically, we might imagine that our scanning electron microscope tip was in fact two tips placed very close together, and these tips have slowly moved apart from one another. While a charge measurement around both internal boundary circles must still give zero, a measurement about either internal boundary separately could give a nonzero value, indicating the presence of a topological defect.  If some specific charge $\sigma$ is detected around one boundary circle, then the opposite charge $\overline \sigma$ must be detected around the other circle.  Thus, we have created a particle-antiparticle pair.

We can describe this process by the solid volume given in Figure~\ref{fig:paircreationsolid}, in which successive horizontal slices from bottom to top represent the changing topology of the disk over time. This solid volume is formed from an upright solid cylinder, with a bent solid cylinder subtracted.
\def\yrad{0.6}
\def\ysep{2}
\begin{figure}[tb]
\centering
$\begin{aligned}
\begin{tikzpicture}[thick, scale=0.8]
\draw [black!20, fill=black!7] (0,0) arc (0:360:2 and \yrad);
\draw (0,0) arc (0:-180:2 and \yrad);
\draw (0,0) to (0,2*\ysep);
\draw (-4,0) to (-4,2*\ysep);
\draw [black!20] (-1.75,2*\ysep)
    to [out=down, in=right, in looseness=0.5] (-2,1.2*\ysep)
    to [out=left, in=down, out looseness=0.5] (-2.25,2*\ysep)
    to (-3.25,2*\ysep)
    to [out=down, in=left] (-2,0.7*\ysep) 
% node [below] {$H$} 
to [out=right, in=down] (-0.75,2*\ysep);
\draw [white] (-3.25,2*\ysep) to (-2.25,2*\ysep);
\draw [fill=black, fill opacity=0.07] (0,2*\ysep) arc (0:360:2 and \yrad);
\draw [fill=white, fill opacity=1] (-2.25,2*\ysep) arc (360:0:0.5 and 0.25);
\draw [fill=white] (-0.75,2*\ysep) arc (360:0:0.5 and 0.25);
%\node at (-3.55,2*\ysep) {$Y$};
%\node [black!30, anchor=180, inner sep=0pt] at (-3.8,0) {$D^2 _{{}}$};
%\node [left] at (-4,\ysep) {\makebox[0pt][r]{${S^1} \!\times I$}};
\end{tikzpicture}
\end{aligned}
\hspace{3cm}
\begin{aligned}
\begin{tikzpicture}[thick, scale=0.8]
\draw [fill=black!7] (0,2*\ysep) arc (0:360:2 and \yrad);
\draw [fill=black!7] (0,\ysep) arc (0:360:2 and \yrad);
\draw [fill=black!7] (0,0) arc (0:360:2 and \yrad);
\draw [fill=white] (-2,\ysep) circle [x radius=0.9, y radius=0.3];
\draw [fill=white] (-0.75,2*\ysep) arc (360:0:0.5 and 0.25);
\draw [fill=white] (-2.25,2*\ysep) arc (360:0:0.5 and 0.25);
\node [rotate=90] at (-2,0.5*\ysep) {$\leadsto$};
\node [rotate=90] at (-2,1.5*\ysep) {$\leadsto$};
\end{tikzpicture}
\end{aligned}$
\caption{Spacetime history of pair creation in a 3d TQFT}
\label{fig:paircreationsolid}
\end{figure}
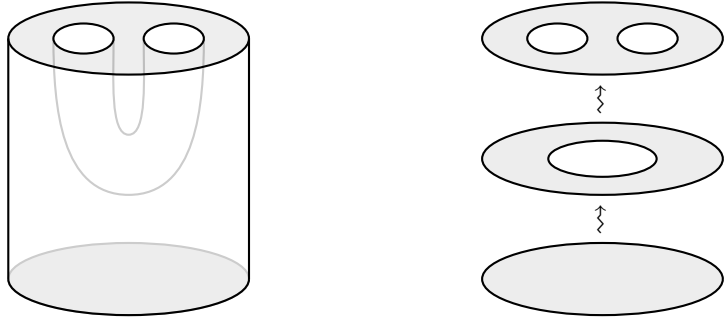

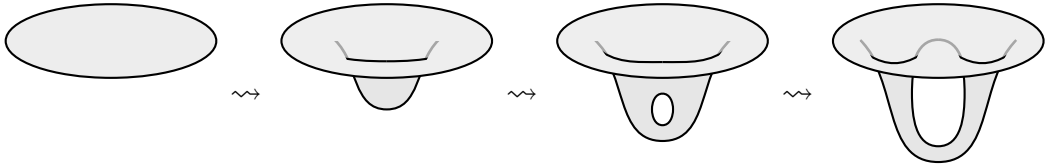
\begin{figure}
\centering
\def\darkfill{black!10}
\def\yrad{0.7}
$\begin{aligned}
\begin{tikzpicture}[scale=0.7, thick]
\path [use as bounding box] (-4,3.7) rectangle (0,0.5);
\draw [fill=black!7] (0,3) arc (0:360:2 and \yrad);
\end{tikzpicture}
\end{aligned}
\,\leadsto\,
\begin{aligned}
\begin{tikzpicture}[scale=0.7, thick]
\path [use as bounding box] (-4,3.7) rectangle (0,0.5);
\def\a{1}
\def\b{2.7}
\def\c{10}
\path [name path=2a] (-2-\a,\b) to [out=-\c, in =-180] (-2,\b-0.085);
\path [name path=2b] (-2,\b-0.085) to [out=0, in =-180+\c] (-2+\a,\b);
\path [name path=1] (-1,3)
    to [out=-135, in=right] (-2,1.7)
    to [out=left, in=-45] (-3,3);
\path [name intersections={of=1 and 2a,by=x}];
\path [name intersections={of=1 and 2b,by=y}];
\begin{scope}
\path [clip] (-5,5) to (x) to (y) to (1,5) to (-5,5);
\clip (-1,3)
    to [out=-135, in=right] (-2,1.7)
    to [out=left, in=-45] (-3,3);
\draw [black!30, line width=2pt] (-1,3)
    to [out=-135, in=right] (-2,1.7)
    to [out=left, in=-45] (-3,3);
\end{scope}
\begin{scope}
\clip (-1,3)
    to [out=-135, in=right] (-2,1.7)
    to [out=left, in=-45] (-3,3);
\draw [name path=2a] (-2-\a,\b) to [out=-\c, in =-180] (-2,\b-0.085);
\draw [name path=2b] (-2,\b-0.085) to [out=0, in =-180+\c] (-2+\a,\b);
\end{scope}
\begin{scope}
\clip (-5,-5) to (-5,5) to (5,5) to (5,-5) to (-5,-5) to (0,3) arc (0:360:2 and \yrad);
\draw [black, fill=\darkfill] (-1,3)
    to [out=-135, in=right] (-2,1.7)
    to [out=left, in=-45] (-3,3);
\end{scope}
\draw [fill=black, fill opacity=0.07] (0,3) arc (0:360:2 and \yrad);
\end{tikzpicture}
\end{aligned}
\,\leadsto\,
\begin{aligned}
\begin{tikzpicture}[scale=0.7, thick]
\path [use as bounding box] (-4,3.7) rectangle (0,0.5);
\def\a{1.3}
\def\b{2.9}
\def\c{40}
\def\d{0.3}
\path [name path=2a] (-2-\a,\b) to [out=-\c, in =-180] (-2,\b-0.3);
\path [name path=2b] (-2,\b-0.3) to [out=0, in =-180+\c] (-2+\a,\b);
\path [name path=1] (-1+\d,3)
    to [out=-135, in=right] (-2,1.7)
    to [out=left, in=-45] (-3-\d,3);
\path [name intersections={of=1 and 2a,by=x}];
\path [name intersections={of=1 and 2b,by=y}];
\begin{scope}
\path [clip] (-5,5) to (x) to (y) to (1,5) to (-5,5);
\clip (-1+\d,3)
    to [out=-135, in=right] (-2,1.7)
    to [out=left, in=-45] (-3-\d,3);
\draw [black!30, line width=2pt] (-1+\d,3)
    to [out=-135, in=right] (-2,1.7)
    to [out=left, in=-45] (-3-\d,3);
\end{scope}
\begin{scope}
\clip (-1+\d,3)
    to [out=-135, in=right] (-2,1.7)
    to [out=left, in=-45] (-3-\d,3);
\draw (-2-\a,\b) to [out=-\c, in =-180] (-2,\b-0.3);
\draw (-2,\b-0.3) to [out=0, in =-180+\c] (-2+\a,\b);
\end{scope}
\begin{scope}
\clip (-5,-5) to (-5,5) to (5,5) to (5,-5) to (-5,-5) to (0,3) arc (0:360:2 and \yrad);
\draw [black, fill=\darkfill] (-1+\d,3)
    to [out=-135, in=right] (-2,1.1)
    to [out=left, in=-45] (-3-\d,3);
\end{scope}
\draw [fill=black, fill opacity=0.07] (0,3) arc (0:360:2 and \yrad);
\draw [fill=white, fill opacity=1] (-2,2.0)
    to [out=right, in=up] (-1.8,1.7)
    to [out=down, in=right] (-2,1.4)
    to [out=left, in=down] (-2.2,1.7)
    to [out=up, in=left] (-2,2.0);
\end{tikzpicture}
\end{aligned}
\,\leadsto\,
\begin{aligned}
\begin{tikzpicture}[scale=0.7, thick]
\path [use as bounding box] (-4,3.7) rectangle (0,0.5);
\def\a{1.5}
\def\b{2.9}
\def\c{40}
\def\d{0.5}
\def\e{0.4}
\def\bottom{0.7}
\path [name path=1a] (-2,\bottom) to [out=left, in=-45] (-3-\d,3);
\path [name path=1b] (-1+\d,3) to [out=-135, in=right] (-2,\bottom);
\path [name path=L] (-5,2.7) to (1,2.7);
\path [name path=Ha] (-2,1) 
    to [out=left, in=down] (-2.5,2)
    to [out=up, in=left] (-2,3);
\path [name path=Hb] (-2,3)
    to [out=right, in=up] (-1.5,2)
    to [out=down, in=right] (-2,1);
\path [name intersections={of=L and 1a,by=w}];
\path [name intersections={of=L and Ha,by=x}];
\path [name intersections={of=L and Hb,by=y}];
\path [name intersections={of=L and 1b,by=z}];
\begin{scope}
\clip (-5,-5) to (5,-5) to (5,5) to (-5,5) to (-5,-5) to (-2,3)
    to [out=right, in=up] (-1.5,2)
    to [out=down, in=right] (-2,1)
    to [out=left, in=down] (-2.5,2)
    to [out=up, in=left] (-2,3);
\path [clip] (-5,5) to (w) to (-2.8,2.5) to (x) to (y) to (-1.1,2.5) to (z) to (1,5) to (-5,5);
\clip (1,5) to (-1+\d,3)
    to [out=-135, in=right] (-2,\bottom)
    to [out=left, in=-45] (-3-\d,3) to (-5,5);
\draw [black!30, line width=2pt] (-1+\d,3)
    to [out=-135, in=right] (-2,\bottom)
    to [out=left, in=-45] (-3-\d,3);
\draw [black!30, line width=2pt] (-2,3)
    to [out=right, in=up] (-1.5,2)
    to [out=down, in=right] (-2,1)
    to [out=left, in=down] (-2.5,2)
    to [out=up, in=left] (-2,3);
\end{scope}
\begin{scope}
\clip (-1+\d,3)
    to [out=-135, in=right] (-2,\bottom)
    to (-2,1)
    to [out=right, in=down] (-1.5,2)
    to [out=up, in=right] (-2,3)
    to [out=left, in=up] (-2.5,2)
    to [out=down, in=left] (-2,1)
    to (-2,\bottom)
    to [out=left, in=-45] (-3-\d,3);
\draw [shorten <=-1pt, shorten >=-1pt] (w) to [out=-30, in=-150] (x);
\draw [shorten <=-1pt, shorten >=-1pt] (y) to [out=-30, in=-150] (z);
\end{scope}
\begin{scope}
\clip (-5,-5) to (-5,5) to (5,5) to (5,-5) to (-5,-5) to (0,3) arc (0:360:2 and \yrad);
\draw [black, fill=\darkfill] (-1+\d,3)
    to [out=-135, in=right] (-2,\bottom)
    to [out=left, in=-45] (-3-\d,3);
\draw [fill=white, fill opacity=1] (-2,3)
    to [out=right, in=up] (-1.5,2)
    to [out=down, in=right] (-2,1)
    to [out=left, in=down] (-2.5,2)
    to [out=up, in=left] (-2,3);
\end{scope}
\draw [fill=black, fill opacity=0.07] (0,3) arc (0:360:2 and \yrad);
\end{tikzpicture}
\end{aligned}$
\caption{Formation of a wormhole}
\label{fig:paircreationmovie}
\end{figure}

As a whole, this solid volume can also be interpreted as a \emph{cobordism with corners}. This means it has a privileged lower and upper boundary---given in this case by a disk, and a disk with a handle glued onto it---such that these boundaries themselves  have identical boundaries, given in this case by a single circle. The cobordism can be interpreted as a set of instructions for the initial boundary to evolve over time into the final boundary, in such a way that their own boundaries are preserved, as shown in Figure~\ref{fig:paircreationmovie}. 
This is quite different to our original intuition in terms of horizontal slices through our solid volume, in which internal boundaries changed over time. Now we have a depression that deepens, in which a hole develops and grows over time, giving rise to a handle that connects two regions of space.  However, the union of all these surfaces is still the solid shown in Figure~\ref{fig:paircreationsolid}: we have merely sliced the same piece of spacetime in a different way.

Here we have the key signature of the ER=EPR correspondence in 3d TQFT. A process that we constructed in Figure~\ref{fig:paircreationsolid} to  describe the creation of a pair of particles is seen in Figure~\ref{fig:paircreationmovie} to correspond to the formation of a wormhole between two points of space. Note that this wormhole formation process is fully local, as emphasized by Maldacena and Susskind~\cite{erepr}: it does not permit the creation of a wormhole between distant points of space without disturbing the intervening manifold.

\section{Mathematical analysis}
\label{sec:doublecategories}

Extended TQFTs can be described using either 2-categories \cite{123tqftIII} or double categories~\cite{Morton}.  In the 2-category approach, space is a manifold with boundary at each moment in time, and the boundary does not change with the passage of time.  In the double category approach, the boundary is allowed to change with time.   Translating between these approaches is crucial to understanding how pair creation is related to wormhole formation.  The first approach is clearly a special case of the second, but in fact the second can be reduced to the first. 

\def\yrad{0.6}
\def\ysep{2}
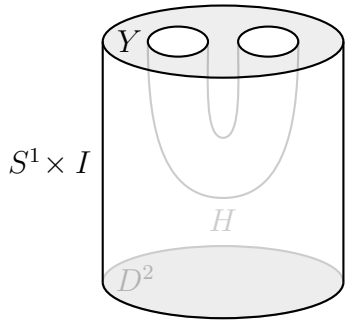
\begin{figure}[h]
\centering
$\begin{aligned}
\begin{tikzpicture}[thick, scale=0.8]
\draw [black!20, fill=black!7] (0,0) arc (0:360:2 and \yrad);
\draw (0,0) arc (0:-180:2 and \yrad);
\draw (0,0) to (0,2*\ysep);
\draw (-4,0) to (-4,2*\ysep);
\draw [black!20] (-1.75,2*\ysep)
    to [out=down, in=right, in looseness=0.5] (-2,1.2*\ysep)
    to [out=left, in=down, out looseness=0.5] (-2.25,2*\ysep)
    to (-3.25,2*\ysep)
    to [out=down, in=left] (-2,0.7*\ysep) 
 node [below] {$H$} 
to [out=right, in=down] (-0.75,2*\ysep);
\draw [white] (-3.25,2*\ysep) to (-2.25,2*\ysep);
\draw [fill=black, fill opacity=0.07] (0,2*\ysep) arc (0:360:2 and \yrad);
\draw [fill=white, fill opacity=1] (-2.25,2*\ysep) arc (360:0:0.5 and 0.25);
\draw [fill=white] (-0.75,2*\ysep) arc (360:0:0.5 and 0.25);
\node at (-3.55,2*\ysep) {$Y$};
\node [black!30, anchor=180, inner sep=0pt] at (-3.8,0) {$D^2 _{{}}$};
\node [left] at (-4,\ysep) {\makebox[0pt][r]{${S^1} \!\times I$}};
\end{tikzpicture}
\end{aligned}$
\caption{Cobordism from disk to disk with handle attached}
\label{fig:paircreationmovie2}
\end{figure}

The formation of a wormhole is described by a cobordism from a disk to a disk with a handle attached. This is a cobordism between two surfaces with boundary, and the boundary does not change with time: it remains $S^1$.  So, we can describe this cobordism in the 2-category approach as a 2-morphism $\alpha \maps D^2 \To Y \circ H$ where $D^2$ is the disk and $Y \circ H$ is the disk with a handle attached.  In more detail, $\alpha$ can be drawn like this:
\begin{equation}
  \xymatrix{
  \emptyset \ar@/^4ex/[rrr]^{Y \circ H}="g1"\ar@/_4ex/[rrr]_{D^2}="g2"&&&S^1
  \ar@{<=}^{\alpha} "g1"+<0ex,-2.5ex>;"g2"+<0ex,2.5ex>
}
\label{eq:wormhole}
\end{equation}
since both $D^2$ and $Y \circ H$ can themselves be seen as cobordisms from the empty set 
to the circle, $S^1$.   This sort of diagram is typical for a 2-morphism in a 2-category.

Pair creation is described by a cobordism from a disk to a disk with two punctures, as shown in Figure \ref{fig:paircreationsolid}. This is again a cobordism between two surfaces with boundary, but now the boundary changes with the passage of time:
\begin{equation}
\label{eq:paircreation}
\begin{aligned}
  \xymatrix{
  S^1 \sqcup S^1 \ar[rr]^{Y}="g1" &&S^1 \\  \\
  \emptyset \ar[rr]_{D^2}="g2" \ar[uu]^H && S^1 \ar[uu]_{S^1 \times I}  
\ar@{<=}^{\alpha} "g1"+<0ex,-4ex>;"g2"+<0ex,4ex>
}
\end{aligned}
\end{equation}
This sort of diagram is called a `square' in a double category.
As before, the disk $D^2$ can be seen as a cobordism from the empty set to a circle.
But the disk with two punctures, called $Y$ here, cannot; instead it is a cobordism from two circles to one. The outside of Figure  \ref{fig:paircreationsolid} is a cylinder $S^1 \times I$, since the outside portion of the boundary does not change with the passage of time: it remains $S^1$.  But the inside portion of the boundary does change, and this change is described by a handle $H$ going from empty set to~$S^1 \sqcup S^1$.   

As already hinted in the previous section, the square for pair creation is just the 2-morphism for the formation of a wormhole viewed in another way.  In fact this is an example of a general construction, due to Ehresmann \cite{Ehresmann}, for building double categories from 2-categories.  In a 2-category we have 2-morphisms of this shape:
\[    
  \xymatrix{
  A \ar@/^4ex/[rrr]^{g}="g1"\ar@/_4ex/[rrr]_{f}="g2"&&&B
  \ar@{<=}^{\alpha} "g1"+<0ex,-2.5ex>;"g2"+<0ex,2.5ex>
}
\]
while in a double category we instead have squares:
\[    
  \xymatrix{
  A' \ar[rr]^{i}="g1" && B \\  \\
  A \ar[rr]_{f}="g2" \ar[uu]^{h} && B' \ar[uu]_{g}  
\ar@{<=}^{\beta} "g1"+<0ex,-4ex>;"g2"+<0ex,4ex>
}
\]
However, if we start with a 2-category, we can build a double category where the above square is defined to consist of the morphisms $f,g,h,i$ together with a 2-morphism of this sort:
\[    
  \xymatrix{
  A \ar@/^4ex/[rrr]^{i \circ h}="g1"\ar@/_4ex/[rrr]_{g \circ f}="g2"&&&B
  \ar@{<=}^{\beta} "g1"+<0ex,-2.5ex>;"g2"+<0ex,2.5ex>
}
\]

In terms of this construction, the square (\ref{eq:paircreation}) for pair creation can be described using the following 2-morphism:
\[
  \xymatrix{
  \emptyset \ar@/^4ex/[rrr]^{Y \circ H}="g1"\ar@/_4ex/[rrr]_{(S^1 \times I) \; \circ \; D^2}="g2"&&&S^1
  \ar@{<=}^{\alpha} "g1"+<0ex,-2.5ex>;"g2"+<0ex,2.5ex>
}
\]
Since $(S^1 \times I)  \circ  D^2$ can be smoothly deformed into $D^2$, this is precisely the 2-morphism for wormhole formation in equation~(\ref{eq:wormhole}).  This gives a rigorous sense which wormhole formation can be reinterpreted as 
pair creation.

We can use this to apply known facts about wormhole formation in the 2-categorical
approach \cite{BK,123tqftIII} to compute what happens in pair creation.  The key is to apply the TQFT, denoted $Z$, to everything in equation (\ref{eq:paircreation}):
\[
  \xymatrix{
  Z(S^1 \sqcup S^1) \ar[rr]^{Z(Y)}="g1" &&Z(S^1) \\  \\
  \emptyset \ar[rr]_{Z(D^2)}="g2" \ar[uu]^{Z(H)} && Z(S^1) \ar[uu]_{Z(S^1 \times I)}  
\ar@{<=}^{Z(\alpha)} "g1"+<0ex,-4ex>;"g2"+<0ex,4ex>
}
\]
The TQFT sends:
\begin{itemize}
\item compact oriented 1-manifolds to 2-vector spaces,
\item 2d cobordisms between compact oriented 1-manifolds to linear functors,
\item 3d cobordisms between 2d cobordisms to natural transformations.
\end{itemize}
In particular, $Z(S^1) = \Mod$ is a modular tensor category whose structure
determines the entire TQFT~\cite{BK,123tqftIII,Wang}.  Standard facts about TQFTs let us evaluate the above
diagram in terms of this, obtaining:
\[
  \xymatrix{
  \Mod \boxtimes \Mod \ar[rr]^{\otimes}="g1" && \Mod \\  \\
  \Vect \ar[rr]_{\iota}="g2" \ar[uu]^{\eta} && \Mod \ar[uu]_{1_\Mod}  
\ar@{<=}^{Z(\alpha)} "g1"+<0ex,-4ex>;"g2"+<0ex,4ex>
}
\]
The elements of this diagram have the following interpretations:
\begin{itemize}
\item $\Vect$ is the category of finite-dimensional vector spaces,
\item $\iota \maps \Vect \to \Mod$ is the linear functor sending $\C$
to $I$, the unit for the tensor product in $\Mod$,
\item  $1_\Mod \maps \Mod \to \Mod$ is the identity functor on $\Mod$,
\item $\Mod \boxtimes \Mod$ is the tensor product of two copies of $\Mod$,
\item $\otimes \maps \Mod \boxtimes \Mod \to \Mod$ is the linear functor
describing the tensor product in the modular tensor category $\Mod$,
\item $\eta \maps \Vect \to \Mod \boxtimes \Mod $ is the linear functor sending $\C$ to $\bigoplus_{\sigma} \sigma \boxtimes \overline{\sigma}$, where we take the sum over simple objects $\sigma$ (that is, types of particles), and $\overline{\sigma}$ is the dual (antiparticle) of $\sigma$.
\end{itemize}
Most important of all is the natural transformation $Z(\alpha)$. To see what particle state it corresponds to, we must evaluate it at $\C \in \Vect$, obtaining a map
\[    Z(\alpha)_\C \maps I \to \bigoplus_{\sigma} \sigma \otimes \overline{\sigma} \]
Here $I$ is the space of
quantum states of the disc, and $\bigoplus_{\sigma} \sigma \otimes \overline{\sigma}$ is the space of quantum states for a wormhole: when 
one end of the wormhole is the particle $\sigma$, the other end is the corresponding
antiparticle. 

Using standard results for 3d TQFT extended to 1-manifolds~\cite{123tqftIII, Walker}, we see that the above map is given as follows:
\tikzset{arrow/.style={decoration={
    markings,
    mark=at position #1 with \arrow{>}},
    postaction=decorate}
}
\tikzset{reverse arrow/.style={decoration={
    markings,
    mark=at position #1 with \arrow{<}},
    postaction=decorate}
}
\begin{align}
Z(\alpha)_\C \; = \;
\sum _\sigma
\,\,\,\frac{d_\sigma}{D}\hspace{-4pt}
\begin{aligned}
\begin{tikzpicture}[scale=0.5]
\draw [reverse arrow={0.22}, reverse arrow={0.79}] (0,1) node [above] {$\sigma$} to (1,-1) to (2,1) node [above] {$\overline{\sigma}$};
\end{tikzpicture}
\end{aligned}
\label{pair.creation}
\end{align}
Here $d_\sigma$ is the quantum dimension of $\sigma$, and $D = \sqrt{\sum_\sigma d_\sigma^2}$, while the Feynman diagram represents the pair creation process $I \to \sigma \otimes \overline{\sigma}$ in the modular tensor category. In short, equation (\ref{pair.creation}) describes the creation of a superposition of particle-antiparticle pairs, with the amplitude for each pair proportional to the quantum dimension $d_\sigma$. 

We can already see the relation to fake entanglement.  The space of states for a wormhole is just a subspace of that for an arbitrary pair of particles:
\[
\bigoplus_{\sigma} \sigma \otimes \overline{\sigma}
\,\subseteq\,
\displaystyle\bigoplus _{\sigma} \sigma
\,\otimes\,
\displaystyle\bigoplus _{\tau} \tau
\]
States of two particles that appear entangled can thus turn out to be `fake entangled' when they are constrained to lie in this subspace by virtue of the fact that the particles are opposite ends of the same wormhole.   However, the relation to entanglement becomes clearer if we consider some specific examples. 

In condensed matter physics, Levin and Wen \cite{LW} introduced a class of `string-net models', essentially 2+1-dimensional lattice gauge theories with finite gauge group $G$ where the space of ground states defines a topological quantum field theory.    In this TQFT,  $\Mod$ is the category of finite-dimensional unitary representations of $DG$, a Hopf algebra called the `quantum double' of $G$.  The tensor product is the usual tensor product of representations.  

In this case, $I$ is the trivial representation of $DG$ on $\C$, and we can take $1 \in \C$ to be the unique state of the TQFT on the disk: what one might call the `vacuum'.  For each irreducible representation $\sigma$ we can choose an orthonormal basis $e_{i, \sigma}$ where $i$ ranges over some set $S_\sigma$, and then
\begin{equation}
\label{counit}
\begin{aligned}   
\begin{tikzpicture}[scale=0.5]
\draw (0,1) node [above] {$\sigma$} to (1,-1);
\draw (1,-1) to (2,1) node [above] {$\overline{\sigma}$};
\draw [->] (2,1) to (1.5,0);
\draw [-<] (0,1) to (0.5,0);
\end{tikzpicture}
\end{aligned} \maps 1 \mapsto \sum_i e_{i, \sigma} \otimes e_{i, \sigma}^* 
\end{equation}
where $e_{i,\sigma}^*$ is the dual basis of $\overline{\sigma}$.    It follows that
\begin{equation}
\label{entangled.state}
       Z(\alpha)_\C  (1) = \sum_{\sigma}  \sum_{i \in S_\sigma} 
\frac{d_\sigma}{D} e_{i, \sigma} \otimes e_{i, \sigma}^* .
\end{equation}
We thus see that, up to normalization, $\psi = Z(\alpha)_\C (1)$ is a completely entangled state in the Hilbert space $(\bigoplus _{\sigma} \sigma) \otimes (\bigoplus_\tau \tau)$, according to our definition of `complete entanglement'.  However, thanks to the presence of the wormhole, $\psi$ is not an arbitrary state of two particles: it is constrained to lie in the subspace $\bigoplus_{\sigma} \sigma \otimes \overline{\sigma}$.   Thus, while we have created a particle-antiparticle pair which superficially appears completely entangled, this entanglement is revealed to be `fake' when we notice that the particle and antiparticle are merely opposite ends of the same wormhole.

One may wonder why $\psi$ is not normalized.  This is due to a well-known fact: in the presence of topology change, the time evolution operators in a TQFT are not unitary, only linear \cite{Bartlett}.  However, it is still impossible to `clone a quantum state' in this context~\cite{Baez}, and the usual proof that true entanglement is monogamous still goes through.

In quantum gravity, a different class of examples is important: the Reshetikhin--Turaev and Turaev--Viro models \cite{BW,RT,TV}.  In the former, $\Mod = \Rep(\U_q \g)$ is a suitably truncated category of finite-dimensional unitary representations of a quantum group $\U_q \g$, where $q$ is a root of unity.  The Turaev--Viro model is a `doubled' version where we take $\Mod = \Rep(\U_q \g) \boxtimes \Rep(\U_{q^{-1}} \g)$.   Quantum gravity in signature +++ with positive cosmological constant is widely believed to be described by the Turaev--Viro model where $\g = \mathfrak{sl}_2$ and $q$ is a certain function of the cosmological constant.  (For a review with many references, see \cite{BH}.)

In the Reshetikhin--Turaev and Turaev--Viro models, equation (\ref{entangled.state}) is only approximately correct.  The reason is that to define the modular category $\Mod = \Rep(\U_q \g)$, we work not with all finite-dimensional representations of $\U_q \g$, but only direct sums of highest weight representations where the highest weight lies in a certain set depending on $q$, called the Weyl alcove \cite{Sawin}.  Thus, when we form the tensor product of two representations, we need to discard part of the resulting Hilbert space to obtain an object of $\Mod$.  Since the tensor product in $\Mod$ differs from the naive tensor product of Hilbert spaces in this way, equation (\ref{counit}) holds only for `small' irreducible representations $\sigma$: that is, those where the sum of the highest weights of $\sigma$ and $\overline{\sigma}$ lies in the Weyl alcove.  The terms in (\ref{entangled.state}) corresponding to `large' representations $\sigma$ need to be corrected.  However, the state $\psi$ will typically exhibit fake entanglement, even if it is not \emph{completely} entangled.

Concretely, in the case of 3d quantum gravity, an irreducible object $\sigma$ in \mbox{$\Mod = \Rep(\U_q \mathfrak{sl}_2) \boxtimes \Rep(\U_{q^{-1}} \mathfrak{sl}_2)$} is labelled by a pair of spins, each lying in the Weyl alcove $\{0, \frac{1}{2}, \dots, \frac{k}{2}\}$, where $k$ depends on the cosmological constant.  The sum and difference of these spins are proportional to the mass and spin of the corresponding type of particle.   So, we are saying that when a wormhole is formed, its ends look like a particle of arbitrary mass and spin together with the corresponding antiparticle.  If we restrict attention to `small' values of mass and spin, this particle-antiparticle pair appears to be in a completely entangled state.  At large values there will be deviations from complete entanglement.  Nonetheless, the robust feature remains that this entanglement is `fake', in the precise sense we have explained.  

\section{Conclusions}
\label{sec:conclusions}

In a 3d topological quantum field theory, when a wormhole forms, a
particle-antiparticle pair of any type can appear at the wormhole's ends.  
The pair are created in an entangled state, if we regard their joint state as a vector in the Hilbert space $(\bigoplus _{\sigma} \sigma) \otimes (\bigoplus_\tau \tau)$. However, when we take account of the fact that the two particles are simply two ends of the same wormhole, we realize that one \emph{must} be the antiparticle of the other, and that their joint state must lie in the diagonal subspace $\bigoplus _\sigma \sigma \otimes \overline \sigma$.  The entanglement is then revealed to be fake, in the sense of Section \ref{sec:fakeentanglement}, and monogamy of entanglement does not apply.

We should emphasize that this rigorous version of ER=EPR makes use of special features of 3d topological field theories: for example, that particles can be described as topological defects.    We should not naively extrapolate these ideas to realistic 4-dimensional physics.  However, our treatment applies to the condensed matter physics of thin films whose ground states are effectively described by a 3d TQFT.

\bibliographystyle{plain}

\end{document}